\documentclass[a4paper,10pt,fleqn,
]{article}
\usepackage{amsmath,amssymb,times,cite
}
\usepackage[]{graphicx}

\begin{document}
%
%


\def\todo{{\bf --- below not edited ---}}
\def\more#1{\footnote{\bf more: #1}}
\def\cN{\mathcal{N}}
\def\cP{\mathcal{P}}
\def\cR{\mathcal{R}}

\title{On Cancellations of Ultraviolet Divergences in
Supergravity Amplitudes
\thanks{IPHT-08/050.  Based on a talk given
  by P. Vanhove at the  3rd meeting of  the RTN `` Constituents,  Fundamental Forces
  and  Symmetries of the  Universe'' in  Valencia (Spain),  and Quarks
  2008 at Sergiev Posad (Russia)}
}

\author
{N. E. J. Bjerrum-Bohr$^a$ and 
Pierre Vanhove$^b$\\ \\ 
$^a$ School of Natural Sciences,\\
Institute for Advanced Study,\\
Einstein Drive, Princeton,\\
NJ 08540, USA\\
\texttt{bjbohr@ias.edu}\\ \\ 
$^b$ Institut de Physique Th\'eorique,\\
CEA, IPhT, F-91191 Gif-sur-Yvette, France\\
CNRS, URA 2306, F-91191 Gif-sur-Yvette, France\\
\texttt{pierre.vanhove@cea.fr}}

\maketitle

\begin{abstract}
 Concrete   calculations  have   pointed  out   that   amplitudes  in
 perturbative  gravity  exhibit  unanticipated  cancellations  taming
 their ultraviolet behaviour  independently of supersymmetry. Similar
 ultraviolet  behaviour  of   $\cN=4$  super-Yang-Mills  and  $\cN=8$
 maximal supergravity has explicitly been observed until three loops.
 These cancellations  can  be connected  to  two  manifest
 features  of gravitational theories:  firstly gauge  invariance from
 diffeomorphism   symmetries  and   secondly   that  amplitudes   are
 colourless and exhibits crossing  symmetry. We will give
 a  simple physical  explanation  of the  cancellations exhibited  in
 gravity amplitudes.  We will discuss these two properties in turn as
 well as the  r\^ole of supersymmetry and string  theory dualities in
 the structure of multiloop amplitudes in supergravity.
\end{abstract}

\section{Introduction}
The theoretical construction of unification models for particle
physics has led to remarkable progress in the understanding of
fundamental interactions in Nature. However a complete theory for
gravity is still illusive and it is expected that subtle quantum
gravity effects will play an important role in understanding the
outstanding fundamental problems of modern cosmology and models for
particle physics. Since the discovery of quantum mechanics in the
last century, physicists have been pursuing the construction of a
consistent theory for quantum gravity in order to gain a complete
understanding of quantum gravitational effects at all scales. Field
theories with point-like interactions for gravity in four dimensions
are non-renormalisable because of the dimensionality of the
gravitational coupling constant. No known symmetry has so far been
shown capable of regulating the ultraviolet divergences for such a
theory {\it although} such constructions have not been proven either
to be impossible. Interestingly unique quantum corrections to
gravity can be extracted from treating general relativity as a
point-like effective field theory~\cite{EffGrav}.

\smallskip

String theory provides a consistent framework for quantum gravity
and its supersymmetric extensions. Within this formalism various
gravity amplitudes can be
computed~\cite{Green:1987sp,Polchinski:1998rq}. Expressions for
field theory amplitudes preserving supersymmetry can be derived in
the infinite tension limit ($\alpha'\rightarrow 0$) of  the string.
String  theory  rules for graviton amplitudes that hold at tree
level have  been formulated very elegantly by
Kawai-Lewellen-Tye~\cite{Kawai:1985xq}. Interestingly such rules
also hold in a number of different scenarios~\cite{BernJI,BernKJ}
with various matter contents~\cite{EffKLT}. At one-loop level string
based  rules have  been  formulated for amplitude calculations in
both gauge theory and gravity~\cite{Green:1982sw,StringBased}. The
effect of massive string modes on the low energy effective action of
various compactifications of string theory leads to important
quantum corrections~\cite{Peeters:2000qj} which are relevant for
particle physics unification, moduli
stabilisation~\cite{Becker:2002nn} and
cosmology~\cite{Antoniadis:2002tr}.

String theory combines the effect of a hard ultraviolet momentum
cutoff (determined by the extension of the string while keeping
gauge invariance) and the decoupling of unphysical states thanks to
the modular invariance of its world-sheet theory. Although the theory is perturbatively finite, its
complete degrees of freedom are provided by the non-perturbative
U-duality symmetries~\cite{Hull:1994ys,
Witten:1995ex,Green:2007zzb}.

\smallskip

Power  counting  arguments based on  known  symmetries indicate that
supergravity theories have ultraviolet divergences in four
dimensions and candidates for  explicit counter-terms  at three-loop
order have been constructed~\cite{Howe:1980th,
Kallosh:1980fi,Howe:1983sr,Howe:2002ui,Kallosh:2007}. However
contrary to  the statements from power counting arguments it has
recently been shown by explicit  computation  that one-loop
amplitudes  in  ${\cal  N} = 8$
supergravity~\cite{Bern:1998sv,BjerrumBohr:2006yw,Bern:2007xj,
Benincasa:2007qj,Cachazo:2008dx} can be constructed  from the same
basis of  scalar integrals as ${\cal N}  =   4$  super  Yang-Mills.
Furthermore   divergences  in  four dimensions  in maximal  $\cN=8$
supergravity have  been  shown to  be explicitly    absent     until
three-loop    order     by    direct computation~\cite{Bern:2007hh}.
It  has  been  suggested  that  the absence    of    divergences
might    persist    to   higher    loop
order~\cite{Green:2006gt,Bern:2006kd,Green:2006yu} with the
consequence that  four dimensional $\cN=8$ supergravity could be a
perturbatively finite theory~\cite{BjerrumBohr:2006yw,Bern:2006kd,
Bern:2007hh,Bern:2007xj,Green:2006gt,Green:2006yu}.

\smallskip

The discrepancy between power counting and explicit computation
emphasises the lack of knowledge of the consequences of physical
effects such as (diffeomorphism) gauge invariance in amplitude
calculations~\cite{Bern:2007xj,
ArkaniHamed:2008yf,DZF,BjerrumBohr:2008vc,Bianchi:2008pu,BjerrumBohr:2008,Bern:2008qj}.
This together  with suggestions of possible finiteness of $\cN=8$
supergravity is a  motivation for reconsidering the ultraviolet
behaviour of (super)gravity theories and their relation to string
theory.

\smallskip

This analysis aims to answer the following questions:

\begin{center}
{\sl How can $\cN=8$ supergravity amplitudes be finite?}

{\sl  What r\^ole does string theory symmetries and dualities
play in the possible finiteness?}
\end{center}

\section{One-loop amplitudes in gravity}

A one-loop $n$-graviton amplitude in $D=4-2\epsilon$ dimensions
takes the generic form
\begin{eqnarray}
\label{e:Mn} M_{n ;1}^{(D)} \, &=& \mu^{2\epsilon}\,  \int d^D \ell
\, {\prod_j^{2n} (q_{\mu_j}^{(2n,j)}\ell^{\mu_j}) + \prod_j^{2n-1}
(q_{\mu_j}^{(2n-1,j)}\ell^{\mu_j})
+ \cdots + K \over \ell_1^2\, \cdots \ell_n^2} \\
\nonumber & \equiv& \mu^{2\epsilon}\, \int
d^D \ell \, {\cP_{2n} (\ell)\over \ell_1^2\, \cdots \ell_n^2}
\end{eqnarray}
where $\ell_i^2 =  (\ell - k_1 - \cdots -  k_i)^2$ are the
propagators along the  loop and $q_{\mu_j}^{(i,j)}$ are functions of
external momenta and polarisations.   Because  of  the   two
derivative nature of the cubic gravitational   coupling,  the
numerator $\cP_{2n} (\ell)$ is a polynomial   with    at   most $2n$
power of loop momentum $\ell\equiv\ell_n$.

A one-loop amplitude can be expanded via a succession of
Passarino-Veltman reductions~\cite{Passarino:1978jh} in a linear
basis of $n$-points scalar integrals
\begin{equation}
I^{(D)}_{n}\ =\ \int {d^D\ell\over \ell_1^2\cdots \ell_n^2}
\end{equation}
where $\ell_i^2=(\ell - K_1-\cdots -K_i)^2$ and where
$K_p=k_{i_1}+\cdots +k_{i_p}$ is the sum of momenta flowing into the
corner $p$. A loop amplitude in four dimensions with $2n$ powers of
loop momenta from each vertex can be shown to generically contain
scalar box, triangle and bubble integrals and as well as polynomial
(non-logarithmic) rational terms~\cite{Bern:1994cg}.

\begin{figure}
\centering
\includegraphics[width=10cm]{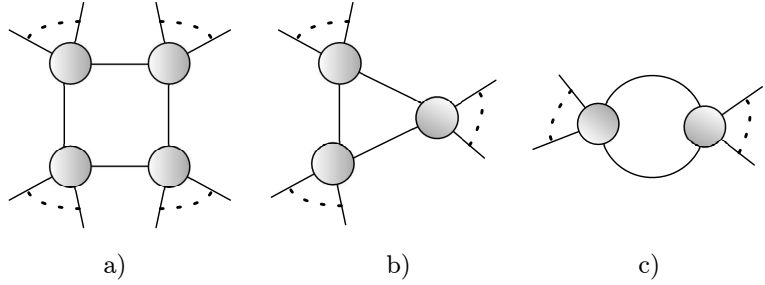}
\caption{\label{fig:integral} Basis of one-loop scalar integrals
given by a) a scalar box, b) scalar triangle and c) a scalar bubble
integral. In  $D=4-2\epsilon$ dimensions these diagrams carry all
the ultraviolet and  infrared divergences of the amplitudes.}
\end{figure}

Explicit evaluation of one-loop gravity amplitudes in
$D=4-2\epsilon$  dimensions  in~\cite{Bern:1998sv,Bern:2005bb,
BjerrumBohr:2005xx,BjerrumBohr:2006yw} show that only scalar box
integrals enter in the decomposition of gravity one-loop amplitudes.
This `only boxes' property  (or the `no triangle hypothesis')
indicates that the \emph{highest} total power of the loop momentum
polynomial in the numerator in the generic one-loop amplitudes has
to be the same as in the ${\cal N} = 4$ super Yang-Mills theory
({\it i.e.} of order $n-4$ and not as na\"\i ve power counting
suggests $2n-8$). For theories with less supersymmetries it was
argued in~\cite{Bern:2007xj} that the highest power of loop momentum
is given by
\begin{equation}\label{e:P}
\cP_{2n}(\ell) \ \sim \ \ell^{2n-\cN-(n-4)},   \quad
\textrm{for}\quad \ell\gg 1
\end{equation}
\noindent This behaviour displays two types of cancellations of loop
momenta,

\begin{itemize}
\item[i)] There is a cancellation  of $\cN$ powers of loop  momenta
due to the effect of the $\cN$ linearised on-shell supersymmetries
(counting
the  number of supersymmetries  in units of  four dimensional
Majorana supercharges). This cancellation is independent of the
number of external  states and the dimension as long as the number
of supersymmetries is preserved.
\item[ii)] There are $n-4$ extra `unexpected'~\cite{Bern:2007xj}
cancellations which depend on the number of external legs.
\end{itemize}

An $\cN=4$ super-Yang-Mills $n$-point one-loop amplitude
contains two kind of contributions. One comes
with at most $n-4$ powers of
loop momentum where $n$ powers of loop momentum come from the
derivatives in the cubic vertices and four cancellations are
due to supersymmetry
\begin{equation}\label{e:threev}
 \int d^D\ell \, {\cP_{n-4}(\ell)\over
\ell_1^2\cdots \ell_n^2}\,,\qquad \ \textrm{with}\  n\geq4
\end{equation}
Another contribution comes with up to $2n-8$  powers of loop momenta
and has many trivial cancellations due to explicit powers of
$\ell_i^2$ in the  numerator. Such contributions lead to
trivial  cancellations such as
\begin{equation}\label{e:fourv}
\int  d^D\ell  {\cP_{n-4}(\ell) \ell_{n+1}^2\cdots \ell_{n+p}^2\over
\ell_1^2\cdots \ell_{n+p}^2}\ =\ \int d^D\ell {\cP_{n-4}(\ell)\over
\ell_1^2\cdots \ell_{n}^2}\,,\qquad \textrm{with}\ n\geq 4
\end{equation}
These types of contributions arise from $\varphi^4$-type of vertices.

Since  the one-loop  $\mathcal{N}=8$ supergravity  amplitude  have the
same maximum number of  loop momenta as $\cN=4$ super-Yang-Mills, they
can be expanded  in the \emph{same} basis of  elementary scalar master
integrals  (in the  dimensional regularisation  scheme).~\footnote{ In
  field theory  we work  using the dimensional  regularisation scheme.
  The momentum cutoff scheme is more natural from the string/M-theory
  point of view,  and will be used later on. Such  a scheme is however
  difficult  to  implement  in  field theory  without  breaking  gauge
  invariance.  In  the  cutoff  regularisation  scheme  the  basis  of
  integrals  could  be  different  in  particular  for  the  integrals
  containing  the  finite  part  of  the  amplitude.}   The  tensorial
structure multiplying  these integrals in  gravity are related  to
the ones   of  the  corresponding   super-Yang-Mills  amplitudes by
the Kawai-Lewellen-Tye
relations~\cite{Kawai:1985xq,Bern:1998ug,Bern:2005bb}.  At  the
level of the effective  action the connection is not immediate
because of the different nature of gauge interactions. There  are no
particular  reasons  for  the higher-derivative corrections  to the
supergravity  effective action  to  be  related directly   via KLT
to  the corresponding contributions   of  the super-Yang-Mills
effective action.   The relation between  the  two theories is
however a useful  guide for the explicit  construction of
higher-derivative gravitational
corrections~\cite{Peeters:2000qj,Dunbar:1999nj}.

\section{Origin of the cancellations}

For a theory with $\cN$ on-shell linearly realised
supersymmetries the integration over the fermionic zero modes leads
to  the cancellation  of $\cN$ powers of loop momenta.

\smallskip

At one-loop level the  extra cancellations of powers of loop momenta
was shown in~\cite{BjerrumBohr:2008vc,BjerrumBohr:2008} to be
accounted for firstly (a) by the summation over all the permutations
of the external legs due to the absence of the concept of colour in
gravity and secondly (b) by the decoupling of longitudinal modes
from the diffeomorphism gauge invariance.

\smallskip

\begin{itemize}

\item[a)]  The absence  of  colour which forces a summation over all the
orderings of the  external legs  leads to  various  cancellations
for on-shell amplitudes.  At  higher-loop order this implies that
one should sum over all planar and non-planar contributions. From
this, four dimensional gravity amplitudes have a better infrared
behaviour  than the  corresponding (coloured)  ordered  QCD
amplitudes \cite{Weinberg}. This gives a  set  of  reduction
formulas needed for the reduction of loop integrals in a basis of
elementary scalar box
integrals~\cite{BjerrumBohr:2008vc,BjerrumBohr:2008}.

\item[b)]  The  diffeomorphism  symmetries  $  \varepsilon_{\mu\nu}\to
  \varepsilon_{\mu\nu}+\partial_\mu v_\nu+ \partial_\nu v_\mu$
and the decoupling of the longitudinal modes in the amplitudes
allows the cancellation of the highest  powers  of   loop  momenta
in the loop integral   by  using unordered integral reduction
formulas~\cite{BjerrumBohr:2008vc,BjerrumBohr:2008}.
\end{itemize}

The                 string                 based
rules~\cite{StringBased,BjerrumBohr:2008vc,BjerrumBohr:2008} give
very  compact and well organised expressions for amplitudes. The
position of  the external  legs are in this formalism labelled by
$\nu_i$ which take values over the range $[0,1]$. The unordered
scalar $n$-point amplitude is given by
\begin{equation}
\mu^{2\epsilon}  \int {d^D\ell\over \pi^{D\over2}} \prod_{i=1}^n {1\over  (\ell-k_1-\cdots
    -k_{i})^2}
=\Gamma\left(D-1\over 2\right)
\Gamma\left(n-{D\over2}\right) \int_0^1 d\nu_1\cdots d\nu_{n-1}\, \delta(\nu_n=1)\, Q_n^{{D\over2}-n}
\end{equation}
where $\mu$ is an infrared regulator and
\begin{equation}\label{defQn}
Q_n=\sum_{1\leq i<j\leq n}\, (k_i\cdot k_j)\,
[(\nu_i-\nu_j)^2-|\nu_i-\nu_j|]
\end{equation}
For a given orderings of the external legs the absolute value in
eq.~(\ref{defQn}) can be lifted. The absolute value forces the
$n$-point integral to break into various regions of analyticity in
the complex energy momentum plane corresponding to the possible
physical orderings of the external legs. By keeping the absolute
value in $Q_n$ and by integrating the $\nu_i$ freely over the range
$[0,1]$ all ordering are included.

For gravitational amplitudes the numerator
$\mathcal{P}(\varepsilon_{ij}, k_i,\ell)$ of the integrand of the
loop amplitude in eq.~(\ref{e:Mn}) depends on the polarisation
vectors $\varepsilon_{ij}$, the external momenta $k_i$ and the loop
momentum $\ell$, with the counting given by eq.~(\ref{e:P})
\begin{equation}
 M^{(D)}_{n;1}= \int {d^D\ell\over  (2\pi)^D} \, {\mathcal{P}(\varepsilon_{ij}, k_i,\ell)\over
   \ell^2\cdots (\ell-k_1-\cdots- k_n)^2}
\end{equation}
Within  the string  based rules,  $\mathcal{P}(\varepsilon_{ij},
k_i,\ell)$  contains  two types of contributions.  One contribution
$\mathcal{P}^{(1)}(\varepsilon_{ij}, k_i,\ell)$ which does {\it not}
involve the the zero mode contribution of  the   bosonic correlators
and one contribution $\mathcal{P}^{(2)}(\varepsilon_{ij}, k_i,\ell)$
which does (defining $\varepsilon_{ij}\equiv(h_i\bar h_j+h_j \bar
h_i)/2$)
\begin{equation}\label{e:zm}
\langle              h_i\cdot\partial_{\nu_i}             X(\nu_i)\,
h_j\cdot\partial_{\nu_j}X(\nu_j)\rangle= (h_i\cdot h_j)\,
[\delta(\nu_i-\nu_j)-1/T]
\end{equation}
This second contribution gives rise to dimension shifted integrals~\cite{BjerrumBohr:2008vc,BjerrumBohr:2008}.

 By expanding the  polarisations of the
external  states  in  a  basis  of  independent  momenta~\cite{BjerrumBohr:2008vc,BjerrumBohr:2008},  the
factor $\mathcal{P}^{(1)}(\epsilon, k_i,\ell)$ can be rewritten as a
homogeneous polynomial of order $2n-\cN$ in  supergravity (or
$n-\cN$ in super-Yang-Mills) in the first  derivative of $Q_n$
 and the ``fermionic
propagator''
$G_F(x)=\textrm{sign}(x)$
\begin{equation}
\label{e:Pex}
 \mathcal{P}^{(1)}(\epsilon,      k_i,\ell)      =      \sum_{r+s=2n-\cN}     c_{r,s}      \,
 \prod_{i=1}^r     \partial_{\nu_i}Q_n     \,     \prod_{\rm {\it s}~pairs~({\it pq})}
 G_F(\nu_p-\nu_q)
\end{equation}
Only   the  contributions   with    $r>   n-\cN/2$  contain triangle
contributions. The coefficients $c_{r,s}$ are functions of the
external momenta for which exact  expressions are  not  needed for
showing  the  absence  of triangles in the amplitude. We remark that
because
\begin{equation}
\partial_{\nu_i}   Q_n  =-2\sum_{j=1}^n   (k_i\cdot  k_j) \,  \nu_j  -
\sum_{j=1}^n (k_i\cdot k_j)\, \textrm{sign}(\nu_i-\nu_j)
\end{equation}
only  the first  derivative  in  $Q_n$ bring  dependence  on the
loop momentum (through its dependence on the $\nu_i$ variables). The
second derivative in $Q_n$ is given by
\begin{equation}
\partial_{\nu_i}\partial_{\nu_j}   Q_n   =2   (k_i\cdot  k_j)   \,
(\delta(\nu_i-\nu_j)-1),\qquad i\neq j
\end{equation}
The first contribution produces  an integral with one less propagator
generating a massive external leg as represented in figure~\ref{fig:integral}
(such contributions  arise from  the reducible contributions  when two
vertex operators are colliding in  string theory~\cite{BjerrumBohr:2008vc}).

In the amplitude for each term in the sum eq.~(\ref{e:Pex}) one can
reduce the number of loop momenta by integration by parts. For the
unordered integrals the boundary  terms   vanish,  but the
integration by  parts generates ultraviolet and infrared finite
contributions given by the dimension shifted scalar integrals

\begin{equation}\begin{split}
I^{(D+2\delta)}_n&=\mu^{2\epsilon}  \int    {d^D\ell   d^{2\delta}
\ell_{\perp}\over \prod_{i=1}^n (\ell- k_1-\cdots
-k_i)^2+\ell_\perp^2}\cr
 & = \Gamma\left(D+2\delta-1\over2\right)\int_0^\infty {dT\over T}\, T^{n+\delta-D/2}
\int_0^1d\nu_1\cdots d\nu_{n-1}\delta(\nu_n-1)\, e^{-TQ_n}
\end{split}\end{equation}
These contributions  have only  four dimensional external  momenta
and combine with the polarisation dependent contributions from
$\mathcal{P}^{(2)}(\varepsilon_{ij},k_i,\nu_i)$ in eq.~(\ref{e:zm})
into gauge invariant expressions.  There were shown
in~\cite{BjerrumBohr:2008vc,BjerrumBohr:2008}  to
 cancel from the physical amplitude.

The  origin of the absence of triangles and bubbles  can be  traced
back to the gauge invariance of the amplitude (the ` cancelled propagator argument'~\cite{Green:1987sp,Polchinski:1998rq}), where the
longitudinal polarisation decouple. Substituting $h_j$ with $i\,
k_j$ in eq.~(\ref{e:zm}) one has
\begin{equation}
  \langle
  h_i\cdot   \partial_{\nu_i}X(\nu_i)\,  \partial_{\nu_j}(e^{ik_j\cdot
    X})\rangle\big|_{\rm linear~in~{\it k_j}}
\end{equation}
which gives a total derivative that cancels against the one
generated by integrating by parts in
$\mathcal{P}^{(1)}(\varepsilon_{ij},k_i,\nu_i)$.

\section{Consequences for the ultraviolet properties of gravity amplitudes}

\subsection{One-loop amplitudes}

The  behaviour  in  eq.~(\ref{e:P})  indicates that  the
$n$-graviton one-loop amplitude has ultraviolet divergences in
dimensions
\begin{equation}\label{e:Dn}
D^{\rm 1-loop}\geq D^{\rm 1-loop}_c\ =\ \cN+n-4
\end{equation}
For more  than   four   gravitons   the  critical   dimension in
eq.~(\ref{e:Dn}) indicates that one-loop gravity amplitudes are more
converging that na\"\i vely expected from supersymmetric
cancellations alone. This leads to the critical dimension for
ultraviolet divergences $D^{\rm susy}_c=\cN$.

\smallskip
For  $\cN=8$ supergravity  the one-loop gravity amplitude would be
finite in eight  dimensions for at least five gravitons and finite
in ten dimensions for at  least seven gravitons. For $\cN=4$
supergravity one-loop amplitudes are finite in four dimensions for
at least five gravitons.

\subsection{Higher-loop amplitudes}

At $L$ loop order linearised on-shell supersymmetry implies that the
critical dimension  for ultraviolet divergences in the four-graviton
amplitude is given by
\begin{equation}
D\ \geq\  2+{c_\cN\over L}
\end{equation}
This implies that supergravity theories are finite in two dimensions and
that they are not finite in four dimensions.
The loop order for the appearance of the first logarithmic divergence
is determined by the value of $6\leq c_\cN\leq 18$ depending on the
implementation  of  the linearised  on-shell  supersymmetries determining
the  mass dimension of  the first possible  counter-term to
the supergravity theory~\cite{Howe:1980th,Kallosh:1980fi,
Howe:1983sr,Howe:2002ui,Green:2006yu}.

A $L$ loop $n$-graviton amplitude has mass dimension
\begin{equation}
[M_{n;L}^{(D)}]\ =\  {\rm mass}^{(D-2)L+2}
\end{equation}
The low energy limit of the four-graviton amplitude at $L$ loops reads
\begin{equation}\label{e:M4}
[M_{4;L}^{(D)}] = {\rm mass}^{(D-2)L-(6+2\beta_L)}\, \partial^{2\beta_L}\!R^4
\end{equation}
where we have used  that $\cN=8$ supergravity four-graviton amplitudes
have a  factor of $R^4$  and allowed $2\beta_L$ powers  of derivatives
distributed on the four Riemann tensors. 
The behaviour in eq.~(\ref{e:M4})  indicates that the amplitude
should be expanded in a basis of $L$-loop integrals with the mass
dimension\footnote{This basis contains  planar and
  non-planar  contributions and  some integrals  will  have numerators
  with momentum factors~\cite{Bern:2006kd}.}
\begin{equation}\label{e:I4}
[I_{4;L}^{(D)}]\ =\  {\rm mass}^{(D-2)L-(6+2\beta_L)}
\end{equation}
and a critical dimension for ultraviolet divergences given by
\begin{equation}\label{e:Dc1}
D\ \geq\  2+ {6+2\beta_L\over L}
\end{equation}

When $\beta_L=L$ at  each loop order two extra  powers of the
external momenta  are  factorised and  the  critical  dimension for
ultraviolet divergences is given by~\cite{Green:2006gt,Green:2006yu}
\begin{equation}
\label{e:Dc}
D\ \geq\  D_c=4+{6\over L}
\end{equation}
This is the same critical dimension as $\cN=4$ super-Yang-Mills and
would imply finiteness in four dimensions if valid at all loop
orders. 

As soon as $\beta_L$ is bounded after some loop order, the
relevant critical dimension is given by~(\ref{e:Dc1}) and the theory
will have an ultraviolet divergence in four dimensions.
The  pure  spinor formalism  gives  a counting  of
  supersymmetric   zero  modes  valid   in  all
  dimensions between  $4\leq D\leq 11$ where  $\cN=8$ supergravity can
  be defined.  This construction implies~\cite{Berkovits:2006vc}  that $\beta_L=12$ for $L\geq
  6$  and a  critical dimension  for ultraviolet  divergence  given by
  $D\geq 2+18/L$ according~(\ref{e:Dc1})  which indicates that in four
  dimensions     the    first     divergence     would    occur     at
  nine-loop~\cite{Green:2006yu} with a counter-term given by the expression~(\ref{e:PSDterm}).
 
\smallskip

The rule $\beta_L=L$ is  the optimal one  for finiteness  in four
dimensions. If $\beta_L$ grows slower than $L$ the theory will not
be finite  in four  dimensions.  For instance $\cN=4$ supergravity
is expected  to  satisfy  the  rule  $\beta_L=L/2$ and  have  a
critical dimension for  ultraviolet divergences given by $D\ \geq\
3+6/L$, and a first  divergence at  $L=6$ loops  in four dimensions.
If $\beta_L$ grows faster than  $L$, the theory would be  too
finite.  For instance the  $L$  loop  (planar  and   non planar)
ladder  diagrams  of  the four-graviton amplitudes are all
two-particle cut constructible and are given by  scalar $\varphi^3$
diagrams with a  prefactor satisfying the rule $\beta_L=2(L-1)$.
These diagrams  are ultraviolet finite for $D\leq  6$ which means
that the  leading ultraviolet  divergences of $\cN=8$ amplitudes are
not contained in these ladder diagrams.

\smallskip

When  the  rule  $\beta_L=L$  is  satisfied at  each  loop  order
the four-graviton amplitudes get a new ultraviolet primitive
divergence of order $\Lambda^{D-4}$  which is typical  of
``effective'' interactions of  the  type  of $\cN=4$
super-Yang-Mills. Amplitudes satisfying this rule should be
expandable in the same basis of integrals as $\cN=4$
super-Yang-Mills, but since gravity has no colour, one must include
the planar and non-planar diagrams.

\smallskip

The absence of triangles and bubbles at one-loop order implies via
general factorisation theorems that higher-loop amplitudes cannot
contain diagrams factorisable in one-loop amplitudes containing
triangles or bubbles.  This constraint affects the structure of the
higher loop amplitude~\cite{Bern:2006kd} but is not a sufficient
condition for perturbative finiteness which requires further subtle
cancellations between triangle free
contributions~\cite{Bern:2007hh}.

\smallskip

\section{Relation to string/M-theory and string theory dualities}

In  the previous  section we only discussed  the effects  of on-shell
supersymmetries,  diffeomorphism  invariance and    the
absence  of  colours. In this section we will discuss
the  r\^ole of string theory dualities.

\smallskip

The  rule $\beta_g=g$ implies  that the  $\partial^{2g}R^4$  couplings to the ten-dimensional  string
theory effective action receive perturbative contributions up to
genus $g$~\cite{Green:1997as,
Green:1999pu,Green:2005ba,Green:2006gt,grv:L=2}. This rule has been
directly shown up to genus six using the pure spinor
formalism~\cite{Berkovits:2004px,Berkovits:2006vc}.

\smallskip

The  eleven   dimensional  incarnation  of   $\cN=8$  supergravity
is non-renormalisable with a two-loop logarithmic divergence as
indicated by  the  formula~(\ref{e:Dc}).   The  associated
counter-term  is  the dimension twenty operator $\partial^{12}R^4$
(see as  well
ref.~\cite{Deser:1998jz}). After having reduced the gravity
integrals to the scalar integral basis, one can regulate the
integrals with a local ultraviolet cutoff $\Lambda$ in eleven
dimensions without breaking gauge
invariance~\cite{Green:1997as,Green:1999pu,Green:2005ba,grv:L=2}.
This  is  more  suitable  for  extracting  the  contributions  to
the effective action.  The cutoff should be determined by the
microscopic degree of freedom  of M-theory and  is related  to the
tension of  the M2-brane $T_{M2}\sim 1/\ell_P^3$ or  the M5-brane
$T_{M5}\sim 1/\ell_P^6$.  The $\cN=8$ supersymmetric  cancellations
of loop momenta  in the one-loop amplitudes   imply   that   the
highest  power   of   the   one-loop sub-divergences is  given by
$\Lambda^3$  and more generally  one gets the  following infinite
series of counter-terms to the four-point M-theory effective
action~\cite{Green:1997as, Green:1999pu,Green:2005ba,grv:L=2}
\begin{equation}\label{e:Mtheory}
S_{\rm {\it M}-theory} \ = \ {1\over   \ell_P^9}\,\int  d^{11}x\,
\left[\cR_{(11)}+
  \sum_{k\geq 0} c_k \,\ell_P^{6k+6} \,\nabla^{6k} R^4\right]
\end{equation}
given by powers of covariant derivatives distributed on the Riemann tensors.
This   is  precisely   the   series   of higher-derivative
corrections to the M-theory  effective action that is selected by
the strong coupling limit of  string theory~\cite{Russo:1997mk}. The
coefficients are constrained by the  microscopic degrees of  freedom
of  M-theory and  its  duality symmetries,  and have  been
determined up to order $k=2$ in~\cite{Green:1997as,
Green:1999pu,Green:2005ba,grv:L=2}.
Using the  renormalisation scheme where the value  of the
counter-term is fixed by  the relation between multiloop amplitudes
in M-theory and string theory  and its  duality symmetries, one
finds that  the $R^4$ counter-term  is  fixed  by  the   value  of
the  type~II  genus  one four-graviton  contribution, the  $\partial^6R^4$
by  the genus  two  and the $\partial^{12}R^4$ by the genus three
contribution~\cite{Green:1997as, Green:1999pu,Green:2005ba,grv:L=2}.
\smallskip

We  consider  a  Kaluza-Klein  expansion  of  the  eleven
dimensional multiloop amplitudes  on a  circle of radius
$R_{11}\,\ell_P$.  Using the   M-theory   conjecture
\cite{Witten:1995ex}  which   identifies $R_{11}^3=g_s^2$     and
$\ell_P=\sqrt{R_{11}}\,     \ell_s$    one finds~\cite{Green:2006gt}
that in the string weak coupling limit, where $R_{11}\to 0$,  the
higher-derivative $\partial^{2g}R^4$ couplings to the low-energy expansion
of    type~II   superstring    satisfy   the non-renormalisation
condition $\beta_g=g$ to all orders in perturbation.

\section{Conclusion and discussion}

\smallskip

We have  discussed cancellations that  could be enough for  making
the ungauged  $\cN=8$ supergravity  theory perturbatively  finite.
It is interesting  to   note  that   a  non-renormalisable  theory
with  a dimensionful  coupling constant  still  can have  a
surprisingly  good perturbative  ultraviolet  behaviour.   Ungauged
${\cal  N}   =  8$ supergravity is  unlikely to  lead directly to
relevant phenomenology because of  the absence  of chiral matter.
However since most  of the cancellations  in eq.~(\ref{e:P}) take
place independent  of supersymmetry  it is expected  that
interesting results  can be  obtained in  theories with less
supersymmetry and with more phenomenological relevance.  \medskip

We  have   not  discussed   the  properties  of   gauged
supergravity
theories~\cite{deWit:1982ig,deWit:2007mt,Hull:2002cv,Andrianopoli:2002aq}
which are  phenomenologically more promising  and seem to  enjoy
nice quantum
properties~\cite{Christensen:1980ee,Curtright:1981wv,Stelle:1982ug}.
The  non-Abelian structure  of the  gauging naturally  contain
duality multiplets   under  the   full  U-duality   group   of
supergravities~\cite{deWit:2008ta} which bring along new effects in
the loop amplitude and need a separate analysis.

The  local $SU(8)$  R-symmetry  of the  ungauged  $\cN=8$ allows one only
to consider superfields of at least  mass dimension at least $1/2$ leading
to         possible         counter-terms        starting         from
eight-loops~\cite{Howe:1980th,Kallosh:1980fi}
\begin{equation}
\delta\textrm{L}\ =\ \kappa_{(4)}^{d+12}\,    \int   d^4x   d^{32}\theta\,
\det(E)\, \mathcal{L}(R,T)
\end{equation}
where $\mathcal{L}(R,T)$ is a superspace density of mass dimension
$d$. In the full superspace this density has a least mass dimension
2, since it can only be constructed from superfields of at least
mass dimension 1/2. For instance using the mass dimension 1/2
gravitino superfield $\chi_{ijk}^\alpha$  and the dimension  1
vector superfield $W_{\alpha\beta}^{ij}$  invariants under  the
full $E_7\times SU(8)$ symmetry of the ungauged $\cN=8$
supergravity, one can construct       the      possible eight- or
nine-loop
counter-terms~\cite{Howe:1980th,Kallosh:1980fi,Howe:1983sr,Howe:2002ui,Green:2006yu,Kallosh:2007,Kallosh:2008}
given by the following four-point higher-derivative supersymmetric invariants
\begin{eqnarray}
\delta\textrm{L}&=&\kappa_{(4)}^{14}\!\!\int d^4xd^{32}\theta\,
\det(E)(\chi_{ijk}^\alpha\bar\chi_\alpha^{ijk})^2\sim
\kappa_{(4)}^{14}\!\!\int d^4x  \sqrt{-g}(\nabla^{10}R^4+ \textrm{susy~completion})\\
\label{e:PSDterm}\delta\textrm{L}&=&\kappa_{(4)}^{16}\int
d^4xd^{32}\theta \det(E)(W_{\alpha\beta}^{ij})^4\sim
\kappa_{(4)}^{16} \int d^4x
\sqrt{-g}\,(\nabla^{12}R^4+\textrm{susy~completion})
\end{eqnarray}
given  by  the  supersymmetric   completion  of  powers  of  covariant
derivatives distributed on the Riemann tensors.

 The $SO(8)\times SU(8)$ gauged $\cN=8$ supergravity has an on-shell superspace
formulation~\cite{Howe:1981tp} which leads to the possible
counter-terms~\cite{Howe:1981tp}
\begin{equation}\label{e:ctg}
\delta\textrm{L}=\kappa_{(4)}^{d+12}\int             d^4xd^{32}\theta\,
\det(E)\,\mathcal{L}(R,T)\, f(g)
\end{equation}
with $f(g)$ constructed from the gauge fields. This expression reduces
to the ungauged counter-terms in the limit $g\to0$
with $f(g)\to1$.   But  there  exist  as  well an  infinite  set  of  new
counter-terms  which do not  reduce to  counter-terms of  the ungauged
theory.  As  in the  ungauged case only  superfields of  mass dimension
$1/2$ are invariant under the local $SU(8)$ R-symmetry (the new spin 1
superfield from  the gauging are of  mass dimension 1),  and the first
possible  counter-term can  only arise  at  eight loops.  It is
interesting to understand in more details the structure of the loop
amplitudes in these different versions of $\cN=8$ supergravity.

\medskip
This analysis illuminates the importance of string theory dualities
and symmetries and their r\^ole in the cancellations. These
dualities and symmetries appear to be very important in the possible
ultraviolet finiteness of  $\cN=8$ supergravity together with
physical effects such as diffeomorphism invariance of amplitudes,
although further research is needed to fully clarify these matters
at higher loop order.

\section*{Acknowledgements}

We would like to thank G.  Dvali, and M. B. Green for useful discussions
and B. de Wit for  some comments on the draft. PV would  like to
thank the organizers of the 3rd meeting of the RTN ``Forces
Universes'' in Valencia and the organizers of Quarks 2008 for giving
him the opportunity of presenting these results. The authors would
like to thank the Niels Bohr Institute in Copenhagen for hospitality
and financial support. The research of (NEJBB) was supported by
grant DE-FG0290ER40542 of the US Department of Energy. The research
(PV) was supported in part the RTN contracts MRTN-CT-2004-005104 and
by the ANR grant BLAN06-3-137168.

\end{document}